\documentclass[12pt]{article}
\usepackage{latexsym}
\usepackage{amssymb}
\usepackage[english]{babel}
\usepackage{amsfonts}
\usepackage{graphicx}
\usepackage{amscd}

\newtheorem{definition}{Definition}

\newcommand{\D}{\mathcal{D}}
\newcommand{\A}{\mathcal{A}}

\newcommand{\N}{\mathbb{N}}

\newcommand{\RR}{\mathbb{R}}
\newcommand{\C}{\mathbb{C}}

\newcommand{\Sc}{\mathcal{S}}

\newcommand{\W}{\star_{W}}
\newcommand{\Cs}{C^{*}}
\newcommand{\st}{^{*}-}
\newcommand{\M}{\star_{M}}

\newcommand{\xs}{x(\sigma)}

\begin{document}
\begin{flushright}
SISSA 20/2003/fm
\end{flushright}
\begin{center}
{\huge\textbf{Toward the construction of a $C^{*}$-algebra in string field theory  \\[1cm]}}
\small{A. Parodi \footnote{adryparodi@libero.it; aparodi@sissa.it}\\[0.5cm]}
\small{\emph{S.I.S.S.A. - I.S.A.S. International School for Advanced Studies, \\
Via Beirut 2-4, Trieste, Italy}\\[3cm]}
\end{center}
{\center\large\textbf{Abstract\\[0.5cm]}}
In string field theory there is a fundamental object, the algebra of string field states $\A$, that must be understood better from a mathematical point of view. In particular we are interested in finding, if possible, a $\Cs$ structure over it, or possibly over a subalgebra $\mathcal{U}\subset \A$.\\
In this paper we define a $^{*}$ operation on $\A$, and then using a particular description of Witten's star product, Moyal's star product, we find an appropriate pre $\Cs$-algebra $\Sc(\RR^{2n})$ on a finite dimensional manifold, where the finite dimensionality is obtained with a cutoff procedure on the string oscillators number. Then we show that using an inductive limit we obtain a pre $C^{*}$-algebra $\mathfrak S\subset\A$ that can be completed to a $\Cs$-algebra.

\newpage
\tableofcontents
\section{Introduction}
\label{sec:Introduction}
A very important object in String Field Theory (SFT), introduced for the first time by Witten for the bosonic open string in \cite{[7]}, is the algebra of string states $\A$. It includes string states given by the sum of of all possible states arising in first quantization of the string: 
\[\vert \Phi\rangle=\Big( \phi(x)+A_{\mu}(x)a^{\mu}_{-1}+B_{\mu\nu}(x)a^{\mu}_{-1}a^{\nu}_{-1}+\alpha(x)c_{-1}+\beta(x)b_{-1}+\ldots\Big)c_{1}\vert 0\rangle\]
that are used to build the cubic string field action (see \cite{[8]}).\\
But string fields describe also non perturbative solutions like $D$-branes: they were found using a slightly different version of SFT, called Vacuum String Field Theory (VSFT) described in \cite{[19]}. Such solutions are exponentials, whose exponents are quadratic in the string oscillators.\\
\\
String fields are described as functionals over a particular string configuration, given by its Fourier modes $x^{\mu}_{n}$, so we would like to treat those fields as we do for ordinary quantum mechanic where the particle states are functions in $L^{2}(\RR^{m})$ and we have all the appropriate mathematical tools to deal with them. It is clear that for SFT it is much more difficult as we have to deal with infinite dimensional spaces, so we have to find some useful method to treat them.\\
\\
Returning to $D$-branes, a very important tool used in superstring theory to study them is the $K$-theory of spacetime that gives the $D$-branes $RR$-charge (see \cite{[25]}). In bosonic string, $D$-branes are unstable, due to the presence of a tachyonic mode in the spectrum of the brane worldvolume theory, so we expect to have no conserved charges that is, a trivial $K$-theory. 
Nevertheless, as preliminary step in view of mathematical (supersymmetric examples) we are interested in having some informations on the $K$-theory of the algebra $\A$ or of some subalgebra $\mathcal{U}\subset\A$ in order to verify its triviality in the context of SFT.\\
\\
All the above listed motivations bring us to the need of a better mathematical understanding of the string states algebra $\A$. In particular there exists a well known structure that can be very useful for our purposes: it is the $\Cs$-algebra structure. Infact the theory of $\Cs$-algebras allows the extension from finite dimensional to the infinite dimensional case, as we will see, while on the other hand it is intimately related with $K$-theory.\\
In this work we will manage to show that $\A$ is a $\st$algebra, and we will find an appropriate pre $\Cs$-algebra $\mathfrak{S}\subset \A$ which contains many known string state solutions arising in VSFT, like the squeezed state, the butterfly and the stringy version of GMS solitons found in \cite{[20]}.\\
In order to determine such $\mathfrak{S}$ we had to find the right enviroment, infact Witten's product $\W$, as it is defined, is rather obscure from mathematical point of view, but fortunately there are various different descriptions of it. In particular we used the fact that Witten's  product can be also described by Moyal's product (see \cite{[3]}), which is a very well known object.\\
One of the problems we had to face was the fact that, as already said, string field states are defined on an infinite dimensional space whose coordinates are the string modes $x^{\mu}_{n}$ with $1\leq n<\infty$, so in order to remedy this, we considered a cutoff, taking only a finite even number of oscillators and we managed to define an appropriate pre $\Cs$-algebra given by the Schwartz functions $\Sc(\RR^{2n})$. This construction was shown to work for every even finite dimensional space $\RR^{2i}$, so that we have a pre $\Cs$-algebra for every finite even dimensional space. Then taking the inductive limit of the closure of various pre $\Cs$-algebras we extended the construction to the whole infinite dimensional space.\\
\\
The paper is organized as follows.\\
In section 2 we recall some basic definitions about $\Cs$-algebras and give two different definitions of Moyal's product, explaining when they agree.\\
In section 3 there is a brief review of different various approaches to interpret Witten's  product. We have focused mainly on half string approach and, of course, Moyal's approach, as they are intimately related to this work.\\
In section 4 we show that $\A$ is a $\st$algebra, using some different methods.\\
In section 5 we describe the infinite dimensional space $C_{X^{D}}(I)$ of string configurations modelled on a Hilbert manifold. The string fields are defined on it.\\
In section 6 we consider a cutoff so that we can deal with a finite dimensional string configuration space $\RR^{2n}$, and then we analyze the algebra of Schwartz functions $\Sc(\RR^{2n})$ where the composition is given by Moyal's product. Such functional space can be seen as an operator space over a $L^{2}_{\star}(\RR^{2n})$ Hilbert space (where the product between function is Moyal's product instead of ordinary pointwise product) and so it is possible to introduce an operator norm over it such that $\Sc(\RR^{2n})$ becomes a pre $\Cs$-algebra. This construction is valid for each finite even dimensional space $\RR^{2n}$.\\
In section 7 we consider the inductive limits of $\Cs$-algebras obtaining a pre $\Cs$-algebra of functions on the infinite dimensional space $C_{X^{D}}(I)$.\\
Finally in section 8 we comment on the usefulness of the construction.\\
\section{Mathematical preliminaries}
\label{sec2}
In this section we will recall some of the basic constructions needed in this work. It can be useful in order to get the mathematical taste of the problematic.
\subsection{Definition of $\Cs$-algebra}
\label{sec:DefinitionOfCsAlgebra}
In this section we want to recall briefly the basic definition of a $\Cs$-algebra.\\
\begin{definition}\label{def1:ps}
a $\Cs$-algebra $A$ is:\\
\\
$\bullet$ a linear associative (in general noncommutative) algebra over the field $\C$ ;\\
\\
$\bullet$ a Banach algebra, that is a normed space such that:\\
\[\Vert a\Vert\geq 0~;~~\Vert a \Vert=0~~iff~~a=0,~~\forall~a\in A\]
\[\Vert \lambda a\Vert=\vert\lambda\vert\Vert a\Vert,~~\forall~\lambda \in \C\]
\[\Vert a+b\Vert\leq \Vert a\Vert+\Vert b\Vert~~~\forall a,b\in A\]
\[\Vert a\circ b\Vert\leq \Vert a\Vert\Vert b\Vert,~~\forall~a,b\in A\]
The last requirement tells us that the product $\circ$ in the algebra is continuous (clearly $A$ is a complete space with respect to the topology defined by the norm) ;\\
\\
$\bullet$ a $\st$algebra, i.e. there exists an antilinear $^{*}$ operation
\[^{*}:A\longrightarrow A\]
such  that
\[(a+b)^{*}=a^{*}+b^{*};~~(\lambda a)^{*}=\overline{\lambda}a^{*};~~(ab)^{*}=b^{*}a^{*}\]
with the involutionary property
\[a^{**}=(a^{*})^{*}=a,~~\forall~a\in A\]
$\bullet$ it satisfies the $\Cs$ condition:
\[\Vert a^{*}a\Vert=\Vert a \Vert^{2}\]
\end{definition}
If all the conditions but the completeness are fulfilled we have a \emph{pre $\Cs$-algebra} that can be completed to a $\Cs$-algebra adding the missing elements.\\
Actually a generic abstract non commutative $\Cs$-algebra can be described in terms of bounded operators over some appropriate Hilbert spaces, thanks to the GNS theorem (see \cite{[18]}).\\
\\
Infact a very important example of $\Cs$-algebra is given by the space $\mathcal{B}$ of all bounded operators 
\footnote{we recall that a bounded operator on a Banach space $\mathcal{B}an$ is a linear map $A:\mathcal{B}an\rightarrow \mathcal{B}an$ such that
\[\Vert A\Vert:=sup \{\Vert Av\Vert~;v\in\mathcal{B}an~s.t.~\Vert v\Vert=1\}<\infty\]
From such definition follows that 
\[\Vert Av\Vert \leq\Vert A\Vert\Vert v\Vert~~~\forall~v\in \mathcal{B}an\]}
over a Hilbert space $L^{2}(\RR^{m})$ with norm $\Vert~\Vert_{2}$. It can be shown that the operator norm defined by
\begin{equation}
	\Vert \mathcal{O}\Vert:=sup\left\{\Vert \mathcal{O}f\Vert_{2}~~s.t.~~f\in L^{2}(\RR^{m})~;~\Vert f\Vert_{2}\leq 1~;~\mathcal{O}\in\mathcal{B}\right\}
\end{equation}
induces a structure of $\Cs$-algebra on $\mathcal{B}$ (see \cite{[26]}).\\ 
Moreover each operator norm-closed $^{*}$-subalgebra contained in $\mathcal{B}$, is too a $\Cs$-algebra. 
\subsection{Mathematical foundation of Moyal's product}
\label{sec:MoyalProduct}
\subsubsection{Introduction}
\label{sec:Introduction}
Moyal's product is usually applied in the context of quantum mechanic. Infact, given a classical system we can obtain its quantum description by promoting all physical observables (defined by functions on the phase space $M$) to operators:
\[f(p,q)\longrightarrow \widehat{f}(\widehat{p},\widehat{q}).\]
Of course we have a huge change in the nature of observables, in particular the commutative pointwise product defined on the algebra of continuous functions $C(M)$, becomes a noncommutative product in the realm of operators.\\
In the past people tried to understand quantum mechanics remaining in the realm of functions defined on the phase space of a classical system, but changing the product between observables (functions), introducing the so called \emph{Moyal's product} which is a noncommutative, associative product that allows a comparison with the operators approach.\\
\\
There have been two different approaches to the problem which, at first sight seem to be the same, but in general they are not.\\
The first is based on the so called \emph{deformation quantization}, built using a deformation of the usual pointwise product, that aims to obtain a noncommutative, associative product. It can be called \emph{differential Moyal's product}. There is also another definition, based on a different quantization procedure, related to group theoretical considerations that can be called \emph{integral Moyal's product}. Both of them depend on a parameter $\vartheta$ describing the non commutativity (in the case of a quantum mechanical system we have $\vartheta=\hbar$). Taking the limit $\vartheta \rightarrow 0$ it can be shown that for quite general functional and distributional spaces the two definitions agree.\\
For completeness we will recall briefly the two constructions.
\subsubsection{Deformation quantization}
\label{sec:DeformationQuantization}
Let us see the very basics of differential Moyal's product referring to \cite{[4]} and \cite{[5]}.
\paragraph{Poisson Manifolds.}
\label{sec:PoissonManifolds}
First of all we want to generalize to a generic differentiable manifold the Poisson structure that is well defined in the case of $\RR^{2n}$, related to the Poisson brackets.\\
Let us take a differentiable manifold $W$ and the algebra of its $C^{\infty}$ functions $N=C^{\infty}(W,\RR)$.\\
If we consider $W=\RR^{2n}$ the Poisson brackets are  linear operators acting on $N$ 
\[\left\{,\right\}:N\times N\longrightarrow N\]
such that
\[\left\{f,g\right\}=\sum_{i}\Big(\frac{\partial f}{\partial q_{i}}\frac{\partial g}{\partial p_{i}}-\frac{\partial f}{\partial p_{i}}\frac{\partial g}{\partial q_{i}}\Big)\]
Among other properties they satisfy the so called \emph{Jacobi identity}:
\[\left\{\left\{f,g\right\},h\right\}+cyclic=0,~~\forall~f,g,h\in N\]
They can also be described using a skewsymmetric rank 2 controvariant tensor $\Lambda$ acting on the differentials $df=\alpha_{i}dx^{i},dg=\beta_{j}dx^{j}$, by means of the internal product $^{\lrcorner}$.\\
Using $df\wedge dg\in\Lambda^{2}(W)$ we define 
\[\Lambda:\Lambda^{2}(W)\longrightarrow N\]
with
\[df\wedge dg \leadsto \Lambda^{\lrcorner}(df\wedge dg)=\Lambda^{ij}\alpha_{i}\beta_{j}.\]
With this formalism Jacobi identity becomes
\begin{equation}\label{eq60:ps}
\frac{1}{2}[\Lambda,\Lambda]^{\lrcorner}(df\wedge dg\wedge dh)=0
\end{equation}
In the case of $W=\RR^{2n}$, $\Lambda$ is an antisymmetric tensor with entries different from zero only when we have coordinates $x^{i},p^{i}$, while they are zero if we have $x^{i},p^{j}$ with $i\neq j$.\\
\\
On a generic manifold $W$ we do not have this symplectic structure $\Lambda$ and so there are problems to define Poisson brackets.\\
\\
So we will consider particular manifolds, called \emph{Poisson manifolds} that are characterized by the fact that they admit a degree 2 skewsymmetric tensor $\Lambda$ called \emph{Poisson structure} such that the Schouten brackets vanish:
\begin{equation}\label{eq55:ps}
 [\Lambda,\Lambda]=0
\end{equation}
Requirement (\ref{eq55:ps}) is of fundamental importance as the Poisson structure enters  the definition of Poisson brackets and so also in the Jacobi relation. From (\ref{eq60:ps}) we see that if we want them to hold we need exactly the condition (\ref{eq55:ps}).\\
\\
Given a Poisson manifold $W$ we can introduce a particular connection, called \emph{Poisson connection} $\Gamma$ such that
\[\nabla \Lambda=0\]
A flat Poisson manifold is a manifold whose Poisson connection has no curvature.\\
\\
Now we can begin the part on deformation.
\paragraph{ Differential Moyal's product.}
\label{sec:MoyalProduct}
Let us take a flat Poisson manifold and after having considered the classical hamiltonian mechanics on it, we want  to "deform" something in order to obtain the quantum mechanics.\\
As already said one of the main properties which is lost passing from classical to quantum mechanic is the commutativity of the product between observables, while associativity is mantained.\\
So if we want to describe the system using classical functions on phase space we need to change their composition law introducing an associative, noncommutative product that will not be pointwise any more.\\
Let us begin noticing that the Poisson brackets have the structure of a noncommutative product and then see if we can use this fact.\\
So consider a flat Poisson manifold $W$ and bidifferentials operators $P^{r}$ of order $r$:
\[P^{r}:N\times N\longrightarrow N\]
with
\[P^{r}(f,g)=\Lambda^{i_{1}j_{1}}\ldots\Lambda^{i_{r}j_{r}}\nabla_{i_{1}}\ldots\nabla_{i_{r}}f\nabla_{j_{1}}\ldots\nabla_{j_{r}}g\]
Let us consider some cases:\\
$r=0$: by definition we set $P^{0}(f,g)=fg$;\\
$r=1$: as $\nabla_{i}f=\partial_{i}f$ we find Poisson Brackets: $P^{1}(f,g)=\Lambda^{ij}\partial_{i}f\partial_{j}g$;\\
$r=2$: $P^{2}(f,g)=\Lambda^{ij}\Lambda^{pq}~\nabla_{i}\nabla_{p}f~\nabla_{j}\nabla_{q}g$.\\
\\
Let us put all these bidifferentials together using a power expansion, and we expect to find a noncommutative product. So let us consider a function
\[f(z)=\sum^{\infty}_{r=0}\frac{a_{r}}{r!}z^{r}\]
and
\[f(\lambda z)=\sum^{\infty}_{r=0} \frac{a_{r}}{r!}(z\lambda)^{r}.\]
Now we can substitute 
\[z^{r}\rightarrow \Big(\frac{i}{2}\vartheta\Big)^{r}P^{r}\]
where $\vartheta$ is the so called \emph{deformation parameter} that encodes the information about how far we are from commuting coordinates.\\
In this way we obtain an operator given by a formal series. This operator is bidifferential and acts on $N\times N\longrightarrow N$. So we can define a formal deformation of the usual product, given by
\begin{equation}\label{eq61:ps}
f\star_{\lambda}g:=\sum^{\infty}_{r=0}\frac{a_{r}}{r!}\Big(i\frac{\vartheta}{2}\Big)^{r}P^{r}(f,g)\lambda^{r},~~f,g\in N
\end{equation}
It is clearly noncommutative. We have to check it is associative that is
\[(f\star_{\lambda}g)\star_{\lambda}h=f\star_{\lambda}(g\star_{\lambda}h)\]
In \cite{[4]} it is shown that it is associative only if $a_{r}=1~~\forall~r$, that is, if the formal series gives an exponential:
\[f\star_{\lambda}g=\exp\Big(i\frac{\vartheta}{2}\lambda P\Big)(f,g)\]
where $P$ represents the Poisson brackets. This can also be written as (setting $\lambda=1$ and considering $\RR^{2d})$  with coordinates $(x_{1},\ldots,x_{d},x_{d+1},\ldots,x_{2d})$ (that in phase space would be $(q_{1},\ldots,q_{d},p_{1},\ldots,p_{d})$):
\begin{equation}\label{eq63:ps}
f\star g=\exp\Big(i\frac{\vartheta}{2}\Lambda^{ij}\partial_{i}\partial'_{j}\Big)f(x)g(x')\Big\vert_{x=x'}.
\end{equation}
We can immediately notice from (\ref{eq61:ps}) that unless we have particular functions like polynomials or exponentials, it's very difficult to calculate the product of two generic functions, so we need a more treatable definition.\\
Once we have Moyal's product we can define a very important structure which gives an indication of the non commutativity of the coordinates. It is called \emph{Moyal bracket} and is defined by
\begin{equation}	
[x^{i},x^{j}]_{\star}=x^{i}\star x^{j}-x^{j}\star x^{i}=\Lambda^{ij}\vartheta
\end{equation}
\subsubsection{Integral Moyal's product}
\label{sec:IntegralMoyalProduct}
Now we recall briefly the main concepts that bring to the definition of Integral Moyal's product which is realized assigning operators to functions in an appropriate way, and then imposing consistency conditions on the product. The main reference is \cite{[22]}.\\
Let us consider a phase space $X$, with a measure $\mu$ and a Hilbert $\mathcal{H}$ space associated to $X$ in some way. 
\begin{definition}
A \emph{Moyal's quantizer} for $(X,\mu,\mathcal{H})$ is a mapping $\Omega$ of $X$ into the space of bounded self adjoint operators on $\mathcal{H}$ such that $\Omega(X)$ is weakly dense in $\mathcal{L}(\mathcal{H})$, and moreover it has to verify
\begin{equation}
	Tr(\Omega(u))=1
\end{equation}
\begin{equation}
	Tr(\Omega(u)\Omega(v))=\delta(u-v)
\end{equation}
\end{definition}
The quantization of a function $f$ defined on $X$ is defined by
\begin{equation}
	f \rightarrow Q(f):=\int_{X} f(u)\Omega(u)d\mu(u)
\end{equation}
We can also perform the opposite operation: given a linear operator $\mathcal{O}\in \mathcal{L}(\mathcal{H})$ we associate a function by
\begin{equation}
	\mathcal{O}\rightarrow W_{\mathcal{O}}(x):=Tr(\mathcal{O} \Omega(x))
\end{equation}
It can be shown that $Q$ and $W$ are inverse operations.\\
\\
Let us consider now the following Moyal's quantizer: 
\[X=\RR^{2n},~~ u=(q,p),~~ d\mu(u)=(2\pi \vartheta)^{-n}~d^{n}p~ d^{n}q,~~ \mathcal{H}=L^{2}(\RR^{n}),\] 
and the action of Moyal's quantizer on functions $f\in L^{2}(\RR^{n})$, given by the following expression:
\begin{equation}
	(\Omega(q,p)f)(x):=2^{n}\exp\Big(\frac{2i}{\vartheta}p(x-q)\Big)f(2q-x)
\end{equation}
Now if we consider a function $g\in \Sc(\RR^{2n})$ we can determine the action of $Q(g)$ over functions:
\[
	(Q(g)f)(x):=\frac{1}{(2\pi \vartheta)^{n}}\int_{\RR^{2n}} g(q,p)(\Omega(q,p)f)(x)~ d^{n}q~ d^{n}p =
\]
\[= \frac{1}{(2\pi \vartheta)^{n}}\int_{\RR^{2n}} g\Big(\frac{x+y}{2},p\Big) e^{ip(x-y)/\vartheta}f(y)~ d^{n}y~ d^{n}p\]
Now that we have a way to assign an operator to functions we can wonder if there exists a product $\star$ between functions so that the following requirement is fulfilled: 
\begin{equation}\label{eq236:ps}
Q(f\star g)=Q(f)Q(g)
\end{equation} 
where the right hand side is the usual operators product. The answer is yes, and we give the following
\begin{definition}
Given two functions $f,g$, there exists a product, called \emph{integral Moyal's product} $f\star g$ defined so that (\ref{eq236:ps}) holds. It is given by
\begin{equation}\label{eq35:ps}
	(f\star g)(x)=2^{2n}\int d\mu(r) d\mu(r') f(r)g(r') \exp\Big[\frac{2i}{\vartheta}(x\Lambda r+r\Lambda r'+ r'\Lambda x)\Big]
\end{equation}
where $r=(p,q)$ and $\Lambda$ is the Poisson structure.
\end{definition}
\subsubsection{Comparison between the two definitions}
\label{sec:ComparisonBetweenTheTwoDefinitions}
As already said, at first sight the two definitions seem to be equivalent: starting from the differential Moyal product
\begin{equation}\label{eq270:ps}
	(f\star g)(x)=f(x)\exp\Big(i\frac{\vartheta}{2}\Lambda^{ij}\partial_{i}\partial'_{j}\Big)g(x')\Big\vert_{x=x'}
\end{equation}
with $x=(q_{1},\ldots,q_{n},p_{1},\ldots,p_{n})$, we can consider 
\begin{equation}\label{eq272:ps}
  f(x)=\int dk~\tilde{f}(k)~ e^{-ikx}~~;~~
  g(x')=\int dk'~\tilde{g}(k')~ e^{-ik'x'}
\end{equation}
Then plugging (\ref{eq272:ps}) into (\ref{eq270:ps}) and considering the Fourier transform of the result, we find
\begin{equation}\label{eq274:ps}
	\widetilde{(f\star g)}(l)=\int d \xi~ e^{i\frac{\vartheta}{2}\Lambda^{ij}l_{i}\xi_{j}}\widetilde{f}\Big(\frac{1}{2}l+\xi\Big)\widetilde{g}\Big(\frac{1}{2}l-\xi\Big)
\end{equation}
This form is related to (\ref{eq35:ps}) by simply taking the Fourier transforms 
\[\tilde{f}(k)=\int dx ~e^{ikx}f(x)~~;~~\tilde{g}(l)=\int dy ~e^{ily}g(y)\]
then plugging them into (\ref{eq274:ps}) and finally the inverse transform of $\widetilde{(f\star g)}$.\\
\\
From these considerations it seems that differential Moyal product and the integral Moyal product are equivalent. The problem is that some mathematical passages are formal and are well defined only if we work on suitable functional spaces. What we have is that in general the two definitions agree when the non commutativity parameter $\vartheta \rightarrow 0$, although there are particular cases in which the correspondence holds for every values $\vartheta>0$ that is, when the functions that are multiplied are analytic.\\ 
We will not deal with this problem. Very good references on it are \cite{[22]} and \cite{[23]}.
\section{Various different approaches to String Field Theory}
\label{sec:VariousDifferentApproachesToStringFieldTheory}
The initial formulation of String Field Theory, given by Witten in \cite{[7]} is quite formal and it does not actually allow much developements if not suitably reinterpreted. There have been various different approaches to the subject, aimed to understand the rather obscure Witten's noncommutative star product $\W$.\\
Now we will review them briefly in order to have a precise idea of the subject.
\subsection{Original Witten's method}
\label{sec:OriginalWittenApproach}
In its original work Witten defined string field theory by means of a quite abstract algebra $\A$ containing all possible string states $\vert \Phi \rangle$, together with a noncommutative product $\W$.\\
The form of the action is the following
\begin{equation}
	S=\frac{1}{2\alpha'}\int \Phi \W Q\Phi +\frac{g}{3!}\int \Phi\W\Phi\W\Phi
\end{equation}
where $Q$ is the BRST operator.\\
The product has to glue together two half strings in order to describe the string interaction: if we suppose to have two string fields $\vert \Phi \rangle,\vert \Psi \rangle$ then their product 
\[(\vert \Phi \rangle\W \vert \Psi \rangle)=\vert X\rangle\] 
has the following meaning:\\
\setlength{\unitlength}{1mm}
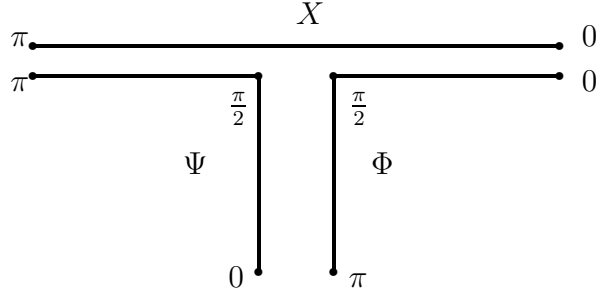
\begin{picture}(150,40)(0,0)
\linethickness{1pt}
\put(30,32){\line(1,0){70}}
\put(30,28){\line(1,0){30}}
\put(60,28){\line(0,-1){26}}
\put(70,28){\line(0,-1){26}}
\put(70,28){\line(1,0){30}}

\put(70,28){\circle*{1}}
\put(60,28){\circle*{1}}
\put(30,28){\circle*{1}}
\put(100,28){\circle*{1}}
\put(30,32){\circle*{1}}
\put(100,32){\circle*{1}}
\put(60,2){\circle*{1}}
\put(70,2){\circle*{1}}

\put(27,32){\makebox{$\pi$}}
\put(103,32){\makebox{$0$}}
\put(65,35){\makebox{$X$}}
\put(103,26){\makebox{$0$}}
\put(27,26){\makebox{$\pi$}}
\put(72,0){\makebox{$\pi$}}
\put(56,0){\makebox{$0$}}
\put(72,23){\makebox{$\frac{\pi}{2}$}}
\put(56,23){\makebox{$\frac{\pi}{2}$}}
\put(50,15){\makebox{$\Psi$}}
\put(75,15){\makebox{$\Phi$}}

\end{picture}
We can give a mathematical meaning to such operation noticing that a string state $\vert \Phi \rangle$ is built using string oscillators $a^{\mu}_{-n}$ acting on the vacuum $\vert 0 \rangle$. If we consider now the string $x(\sigma)$ in terms of Fourier modes, given by
\begin{equation}\label{eq110.1:ps}
x(\sigma)=x_{0}+\sqrt{2}\sum^{\infty}_{n=1}x_{n}\cos(n\sigma),~~~0\leq \sigma \leq\pi
\end{equation}
or also, using the midpoint coordinate $\overline{x}:=x(\frac{\pi}{2})$:
\begin{equation}\label{eq110.2:ps}
x(\sigma)=\overline{x}+\sqrt{2}\sum^{\infty}_{n=1}x_{n}\Big(\cos(n\sigma)-\cos\Big(\frac{n\pi}{2}\Big)\Big)
\end{equation}
then we can express Fourier modes $x_{n}$ in terms of oscillators $a_{-n},a_{n}$ and so we can give a functional form to the string states $\vert \Phi \rangle$. First of all let us define the state $\vert x(\sigma)\rangle$:
\[\vert x(\sigma)\rangle:=\exp\Big(\sum_{n}\Big[-\frac{1}{2}n~ x_{n}x_{n}-x^{2}_{0}+i\sqrt{2n}~a_{n}x_{n}+2i~a_{0}x_{0}+\frac{1}{2}~a_{n}a_{n}\Big]\Big)\vert 0 \rangle\]
with 
\[x_{n}=\frac{i}{\sqrt{2n}}(a_{n}-a_{-n})~~~n\neq 0~;~~~x_{0}=\frac{i}{2}(a_{0}-a^{\dag}_{0})~~\]
so we get
 \[\Phi[x(\sigma)]=\Phi[x_{n}]=\langle x(\sigma)\vert \Phi\rangle.\]
This point will be explained better later on.\\
So using a functional algebra we can rewrite Witten's product in the following formal way:
\begin{eqnarray}\label{eq10:ps}
	(\Phi\W\Psi)[x(\sigma)] & = & \int\prod_{0\leq\sigma\leq\frac{\pi}{2}}\D x_{1}(\sigma)\D x_{2}(\sigma)\times {}
\nonumber	\\
	{} & \times & \prod_{0\leq\sigma\leq\frac{\pi}{2}}\Phi[x_{1}(\sigma)]\Psi[x_{2}(\sigma)]\delta(x_{2}(\sigma)-x_{1}(\pi-\sigma)){}
\end{eqnarray}
But it is not useful for practical calculations.
\subsection{CFT method}
\label{CFTmethod}
If we consider a string field made of all states arising in first quantization given by
\begin{equation}
	\vert \Phi\rangle=\Big( \phi(x)+A_{\mu}(x)a^{\mu}_{-1}+B_{\mu\nu}(x)a^{\mu}_{-1}a^{\nu}_{-1}+\alpha(x)c_{-1}+\beta(x)b_{-1}+\ldots\Big)c_{1}\vert 0\rangle=\]
	\[=\Phi(z=0)\vert 0 \rangle
\end{equation}
then we can use the correspondence between states and operators that we have in CFT (considering a wordsheet approach to the problem), so the interaction between two open strings can be described in terms of OPE of vertex operators (see \cite{[8]} and \cite{[9]}). This is one of the most useful methods if we want to consider the interactions of a finite number of string fields using truncation level techniques (see \cite{[10]}).
\subsection{Algebraic method}
\label{sec:AlgebraicMethod}
There is also a very powerful method developed in \cite{[11]} that uses an algebraic approach based on string oscillators.\\
It starts from the observation that the cubic interaction term can be written using an operator vertex $\langle V_{3}\vert$ (a state in the tensor product $\A^{*}\otimes \A^{*}\otimes \A^{*}$), as follows:
\begin{equation}
	\int \Phi\W\Phi\W\Phi=\langle V_{3}\vert \Phi\otimes\Phi\otimes\Phi\rangle
\end{equation}
and this vertex can be written explicitly by imposing the gluing conditions at algebraic level. The final result is
\begin{equation}
	\langle V_{3}\vert=\delta(p_{(1)}+p_{(2)}+p_{(3)})\langle 0\vert \exp\Big(\frac{1}{2}\alpha^{(r)\mu}_{n}N^{rs}_{nm}\alpha^{(s)\nu}_{m}\Big)
\end{equation}
having considered only bosonic modes and not ghosts.\\
Here the indices $r,s$ take values $1,2,3$ and indicate which string of the three strings involved in the interaction is acted on, while $N^{rs}_{nm}$ with $r,s$ fixed, are the coefficients of an infinite dimensional matrix, and $m,n=1,\ldots \infty$.\\
Now if we want to perform a product between two string fields $\vert \Phi\rangle , \vert \Psi \rangle$ then we can use such vertex:
\[\vert X\rangle =\langle \Phi\vert \langle \Psi\vert \vert V_{3}\rangle.\]
\subsection{Half string method}
\label{sec:HalfStringMethod}
This is based on a technique that manages to break the string oscillators into two distinct sectors, and will turn out to be very useful for our purpose, so we will explain it in more details. The original references are \cite{[7]}, \cite{[12]}, \cite{[13]}.\\
In these works it is shown how the $\W$ product can be interpreted as a matrix product.\\
In order to do it we have to change the description of the string. Let us consider the original function $x(\sigma)$ given by (\ref{eq110.1:ps}) and (\ref{eq110.2:ps}).\\
Then we can break the string into its left half $x^{L\mu}$ and its right half $x^{R\mu}$ defined in the following way:
\begin{equation}\label{eq10.1.ps}
x^{L\mu}(\sigma)=x^{\mu}(\sigma)~~~~~~for~~0\leq\sigma\leq \frac{\pi}{2};
\end{equation}
\begin{equation}\label{eq10.2.ps}
x^{R\mu}(\sigma)=x^{\mu}(\pi-\sigma)~~~~~~for~~0\leq\sigma\leq \frac{\pi}{2};
\end{equation}
Now, both $x^{L\mu}$ and $x^{R\mu}$ can be described in terms of Fourier modes $\{l_{n}\}$ and $\left\{r_{n}\right\}$ respectively and the expansions are given by (see \cite{[3]}, imposing Neumann bounduary conditions at the end of the string: $\partial_{\sigma}x\vert_{0,\pi}=\partial_{\sigma}x^{L}\vert_{0}=\partial_{\sigma}x^{R}\vert_{0}=0$ and Neumann bounduary conditions at the midpoint: $x(\frac{\pi}{2})=x^{L}(\frac{\pi}{2})=x^{R}(\frac{\pi}{2})=\bar{x}$):
\[x^{L}(\sigma)=\overline{x}+\sqrt{2}\sum^{\infty}_{n=1}l_{2n-1}\cos((2n-1)\sigma),~~~0\leq\sigma\leq\frac{\pi}{2},\]
\[x^{R}(\sigma)=\overline{x}+\sqrt{2}\sum^{\infty}_{n=1}r_{2n-1}\cos((2n-1)\sigma),~~~0\leq\sigma\leq\frac{\pi}{2},\]
and there is a relation between the original $x_{n}$ modes of $x^{\mu}(\sigma)$ and $\{l_{n}\},\left\{r_{n}\right\}$. The relations are described by
\begin{equation}
	x_{2n-1}=\frac{1}{2}(l_{2n-1}-r_{2n-1})
\end{equation}
\begin{equation}
	x_{2n\neq0}=\frac{1}{2}\sum^{\infty}_{m=1}T_{2n,2m-1}(l_{2m-1}+r_{2m-1})
\end{equation}
\begin{equation}
	x_{0}=\overline{x}+\frac{1}{4}\sum^{\infty}_{m=1}T_{0,2m-1}(l_{2m-1}+r_{2m-1})
\end{equation}
with
\[T_{2n,2m-1}=\frac{2(-1)^{m+n+1}}{\pi}\Big(\frac{1}{2m-1+2n}+\frac{1}{2m-1-2n}\Big)\]
The inverse relations are
\begin{equation}\label{eq110:ps}
	l_{2m-1}=x_{2m-1}+\sum^{\infty}_{n=1}R_{2m-1,2n}x_{2n}
\end{equation}
\begin{equation}
	\overline{x}=x_{0}+\sqrt{2}\sum^{\infty}_{n=1}(-1)^{n}x_{2n}
\end{equation}
\begin{equation}\label{eq120:ps}
	r_{2m-1}=-x_{2m-1}+\sum^{\infty}_{n=1}R_{2m-1,2n}x_{2n}
\end{equation}
with
\[R_{2m-1,2n}=\frac{4\pi(-1)^{n+m}}{\pi(2m-1)}\Big(\frac{1}{2m-1+2n}-\frac{1}{2m-1-2n}\Big)\]
and $(RT)_{2n-1,2k-1}=\delta_{n,k};~~(TR)_{2m,2l}=\delta_{m,l}$.\\
So a generic string field can be described as 
\[\Phi[x(\sigma)]=\Phi[\left\{x_{n}\right\}]\] 
but also in the split string formalism as 
\[\tilde{\Phi}[x^{L}(\sigma),x^{R}(\sigma)]=\tilde{\Phi}[\left\{l_{2n-1}\right\},\left\{r_{2n-1}\right\}]\]
Now we have given the description of the string fields, it is time to move to Witten's product $\W$ and to understand why it can be seen as a matrix product.\\
\\
Let us recall the functional form of Witten's  product described in (\ref{eq10:ps}).\\
If we use the half string approach the product becomes formally easier, infact we have
\begin{equation}\label{eq15:ps}
	(\tilde{\Phi}\W\tilde{\Psi})[x^{L}_{1}(\sigma),x^{R}_{2}(\sigma)]=\int\D y~ \tilde{\Phi}[x^{L}_{1},y]~\tilde{\Psi}[y,x_{2}^{R}]
\end{equation}
Now we can interpret $x^{L}_{1}$ as the rows of the "matrix" $\Phi[x^{L}_{1},y]$ and $x^{R}_{2}$ as the columns, so the $\W$ becomes a matrix product of infinite dimensional matrices.\\
It can also be written as 
\begin{eqnarray}\label{eq15.1:ps}	
& &(\widetilde{\Phi}\W\widetilde{\Psi})[ \left\{l_{2n-1}\right\},\left\{r_{2n-1}\right\},\overline{x}]= {} 
\nonumber\\
& & {}=\int \prod_{k}dz_{2k-1}\widetilde{\Phi}(\left\{l_{2n-1}\right\},\overline{x},\left\{z_{2n-1}\right\})\widetilde{\Psi}(\left\{z_{2n-1}\right\},\overline{x},\left\{r_{2n-1}\right\}){}
\end{eqnarray}
\subsection{Moyal's method}
\label{sec:MoyalMethod}
Moyal's method is very important in our searching for a $\Cs$-algebraic structure inside $\A$, as it relates Witten's  product to a well known noncommutative, associative product, Moyal's product. Its main definitions and constructions are reviewed in section 2. Here we will focus our attention on \cite{[3]}, the work where the above relation is explained.\\
\\
First of all it is necessary to study another way to perform Moyal's product: so far we have seen the product on the phase space described by the integral kernel (using the integral Moyal product):
\begin{equation}
	L(x,r,r'):=2^{2n}\exp\Big(\frac{2i}{\vartheta}(x\Lambda r+r\Lambda r'+r'\Lambda x)\Big)
\end{equation}
with $x=(q,p)$, and it is the case in which we perform a Fourier transform of all the coordinates on the phase space (see eq. (\ref{eq272:ps})). Now we want to see what happens if we Fourier transform only half of the coordinates and precisely the $p$. Starting from a function $f(p,q)$ we consider
\[f(q,p)=\int dy~ e^{-ipy}\Psi_{f}(q,y)\]
Then we can define two new coordinates
\[l=q+\frac{y}{2}~~;~~ r=q-\frac{y}{2}\]
so that we define a third function $A(l,r)=\Psi_{f}(q,y)$, so that we have
\begin{equation}\label{eq432:ps}
	f(q,p)=\int dy~ e^{-ipy}A\Big(q+\frac{y}{2},q-\frac{y}{2}\Big)
\end{equation}
Now, Moyal's product can be calculated in any variables we want, but it can be shown that if we use the set $(l,r)$ we find a particularly easy form: we consider
\[(f\star g)(q,p)=f(q,p)e^{i\frac{\vartheta}{2}\Lambda^{ij}\partial_{i}\partial'_{j}}g(q',p')\Big\vert_{p=p';q=q'}\]
and then we use the form (\ref{eq432:ps}) for $f,g$.
If we associate $(f\star g)(q,p)\leftrightarrow C(l,r)$ that is
\[(f\star g)(q,p)=\int dy~ e^{-ipy}C\Big(q+\frac{y}{2},q-\frac{y}{2}\Big)\]
then, after some calculations (see \cite{[3]}) we have
\begin{equation}\label{eq80:ps}
C(l,r)=\int d\xi ~A(l,\xi)~B(\xi,r)
\end{equation}
with $f(q,p)\leftrightarrow A(l,r);~g(q,p)\leftrightarrow B(l,r)$; we can easily see that (\ref{eq80:ps}) tells us that Moyal's product can be seen as a sort of matrix product with infinite entries.\\
Summarizing, in Moyal's product using the half Fourier approach, we have the following steps: we start from coordinates on the phase space $(q_{i},p_{i})$ and then obtain $(l_{i},r_{i})$:
\begin{equation}\label{eq100:ps}
\begin{CD}
(q_{i},p_{i})@>\mathcal{F}_{p_{i}}>>(q_{i},y_{i})@>>>\Big(l_{i}=q_{i}+\frac{y_{i}}{2},~ r_{i}=q_{i}-\frac{y_{i}}{2}\Big)
\end{CD}
\end{equation}
and then we have
\begin{equation}\label{eq105:ps}
 C(l_{i},r_{i})=\int d^{D}\xi_{i}~ A(l_{i},\xi_{i})~B(\xi_{i},r_{i})
\end{equation}
But now we can notice the following important fact: (\ref{eq105:ps}) is very similar to Witten's product in split string field theory, described by (\ref{eq15.1:ps}).
So in the case of string field theory we have to perform the transformations in the opposite direction: we have the coordinates $(\left\{l_{2m-1}\right\},\left\{r_{2n-1}\right\})$ corresponding to the previous $(l_{i},r_{i})$. We want to relate them to $(\left\{x_{2n}\right\},\left\{x_{2m-1}\right\})$ which correspond to $(q_{i},y_{i})$ above.
In this case the relation is more complicated than in (\ref{eq100:ps}), infact we have relations (\ref{eq110:ps}) and (\ref{eq120:ps}) with $x_{2m-1}$ the equivalent of $y_{i}/2$ and $R x_{e}:=\sum^{\infty}_{n=1}R_{2m-1,2n}x_{2n}$ the equivalent of $q_{i}$.\\
So now we consider the coordinates $\{Rx_{e},x_{2n-1}\}$ and perform a Fourier transformation of $x_{2n-1}$ to get the conjugate momentum, but instead of using $p_{2n-1}$ we choose a more complicated expression at first sight, that will simplify calculations later on:
\[p_{e}T:=\sum^{\infty}_{n=1}p_{2n}T_{2n,2m-1}\sim p_{i}\]
that is, it plays the role of Moyal's coordinate $p_{i}$ on the phase space.\\
Summarizing we have the following operations:
\begin{equation}\label{eq107:ps}
\begin{CD}
(p_{e}T,Rx_{e})@<\tilde{\mathcal{F}}_{x_{2n-1}}<<(2x_{2n-1},Rx_{e})@<<<(\{l_{2n-1}\},~\{r_{2n-1}\})
\end{CD}
\end{equation}
which are essentially the opposite of (\ref{eq100:ps}).\\
After the Fourier transform of string fields we find (here we use the convention $x_{odd}=\left\{x_{2n-1}\right\}$):
\begin{eqnarray}\label{eq122:ps}
& &\Phi_{\widetilde{M}}(Rx_{e},\overline{x},p_{e}T)= {}
\nonumber\\
& & {}=\int dx_{odd} \exp(-i(p_{e}T 2x_{odd}))\widetilde{\Phi}(Rx_{e}+x_{odd},\overline{x},Rx_{e}-x_{odd}){}
\end{eqnarray}
and this is the expression of the string field theory on the phase space with coordinates
$(Rx_{e},p_{e}T)$.\\
Due to this fact Moyal's brackets of the coordinates $Rx_{e},p_{e}R$ satisfy
\[\Big[(\sum^{\infty}_{n=1}R_{2k-1,2n}x_{2n}^{\mu}),(\sum^{\infty}_{m=1}p^{\nu}_{2m}T_{2m,2l-1})\Big]_{\star}=i\delta_{k,l}\eta^{\mu\nu}\]
but using $(TR)=(RT)=Id$ we can simplify the above expression obtaining the independency for each mode:
\[[x^{\mu}_{2n},p^{\nu}_{2m}]=i\delta_{n,m}\eta^{\mu\nu}\]
At this point we can remove the martices $T,R$ and so we find the string field $\Phi_{M}(\left\{x_{2m}\right\},\overline{x},\left\{p_{2m}\right\})$.
So we get finally the expression of the string field theory on the phase space with coordinates $(\left\{x_{2n}\right\},\left\{p_{2n}\right\})$.\\
The string fields in this final form can be $\W$ multiplied using a Moyal's product described by the coordinates on the phase space, given by equation (\ref{eq35:ps}) with $r=(\left\{x_{2n}\right\},\left\{p_{2n}\right\})$.\\
\\
\textbf{Remark.} There are some calculations that do not depend on the specific functional space we are considering. In this case we can also use the differential form of Moyal product that is
\begin{equation}	\M=\exp\Big(i\frac{\vartheta}{2}\sum^{\infty}_{n=1}\eta^{\mu\nu}\Big(\frac{\overleftarrow{\partial}}{\partial x^{\mu}_{2n}}\frac{\overrightarrow{\partial}}{\partial p^{\nu}_{2n}}-\frac{\overleftarrow{\partial}}{\partial p^{\nu}_{2n}}\frac{{\overrightarrow{\partial}}}{{\partial x^{\mu}_{2n}}}\Big)\Big)
\end{equation}
Sometimes this expression will be more suitable than the integral form to check some properties of the algebra $\A$.
\section{$\A$ as a $\st$algebra}
\label{sec:AAsAStAlgebra}
So far we have simply recalled the main tools involved in SFT. Now we want to begin the search of a $\Cs$-algebra inside $\A$. In this section we will show, using some of the different approaches described previously, that $\A$ is indeed a $\st$algebra.\\
As already said, $\A$ is defined as an abstract noncommutative, associative\footnote{Actually it is not always true: there are some pathological cases in which we have the so called \emph{associativiy anomaly}, for an introduction to the problem see [14],[15]} algebra.\\
Given three fields $\Phi,X,\Psi\in\A$ we have:
\[\Phi\W(X\W\Psi)=(\Phi\W X)\W\Psi\]
and also the distributive property w.r.t. the addition:
\[\Phi\W(X+\Psi)=\Phi\W X+\Phi\W\Psi.\]
We also assume that $\A$ has an identity $I$ such that 
\[I\W\Phi=\Phi\W I=\Phi.\]
Now we want to find an operation $^{*}:\A\longrightarrow \A$ that satisfies the requirements of definition $\ref{def1:ps}$.
\subsection{Witten's approach}
\label{sec:AnalyticApproach}
As already seen, this approach consists in gluing parts of the interacting strings. In particular we want to describe in terms of such prescription the antilinearity:
\begin{equation}
	(\Phi[x_{1}]\W\Psi[x_{2}])^{*}=\Psi^{*}[x_{2}]\W\Phi^{*}[x_{1}]
\end{equation}
So we need some operations that naturally change the order in the $\W$ product. One possibility is given by reversing the orientation of the two strings $x_{1}(\sigma), x_{2}(\sigma)$ so that
\[\sigma\longrightarrow \pi-\sigma\]
so that we define
\begin{equation}\label{eq30:ps}
	(\Phi[x(\sigma)])^{*}:=\overline{\Phi}[x(\pi-\sigma)]
\end{equation}
notice that the complex conjugation is needed in order to fulfill the requirement
\[(\lambda\Phi)^{*}=\overline{\lambda}\Phi^{*}.\]
Let us check if it satisfy also the others:\\
\\
$\bullet$ $\Phi^{**}[x(\sigma)]=(\overline{\Phi}[x(\pi-\sigma)])^{*}=\Phi[x(\sigma)]$;\\
\\
$\bullet$ the check for 
\begin{equation}\label{eq400:ps}
(\Phi[x_{1}(\sigma)]\W\Psi[x_{2}(\sigma)])^{*}=\Psi[x_{2}(\sigma)]^{*}\W\Phi[x_{1}(\sigma)]^{*}
\end{equation}
can be performed graphically considering the changing in the orientations; in this way the following diagram \\
\setlength{\unitlength}{1mm}
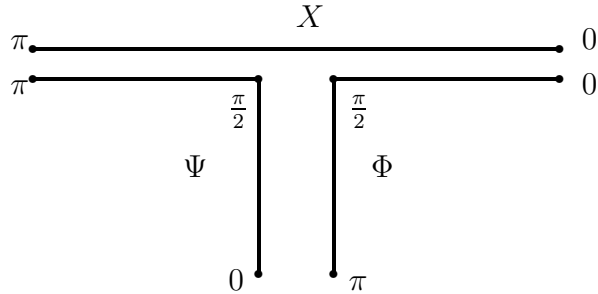
\begin{picture}(150,40)(0,0)
\linethickness{1pt}
\put(30,32){\line(1,0){70}}
\put(30,28){\line(1,0){30}}
\put(60,28){\line(0,-1){26}}
\put(70,28){\line(0,-1){26}}
\put(70,28){\line(1,0){30}}

\put(70,28){\circle*{1}}
\put(60,28){\circle*{1}}
\put(30,28){\circle*{1}}
\put(100,28){\circle*{1}}
\put(30,32){\circle*{1}}
\put(100,32){\circle*{1}}
\put(60,2){\circle*{1}}
\put(70,2){\circle*{1}}

\put(27,32){\makebox{$\pi$}}
\put(103,32){\makebox{$0$}}
\put(65,35){\makebox{$X$}}
\put(103,26){\makebox{$0$}}
\put(27,26){\makebox{$\pi$}}
\put(72,0){\makebox{$\pi$}}
\put(56,0){\makebox{$0$}}
\put(72,23){\makebox{$\frac{\pi}{2}$}}
\put(56,23){\makebox{$\frac{\pi}{2}$}}
\put(50,15){\makebox{$\Psi$}}
\put(75,15){\makebox{$\Phi$}}

\end{picture}
under the $^{*}$ prescription seen above becomes\\
\setlength{\unitlength}{1mm}
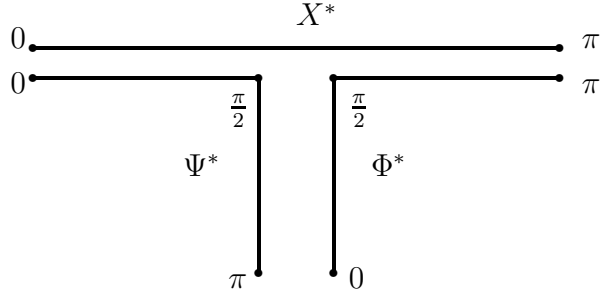
\begin{picture}(150,40)(0,0)
\linethickness{1pt}
\put(30,32){\line(1,0){70}}
\put(30,28){\line(1,0){30}}
\put(60,28){\line(0,-1){26}}
\put(70,28){\line(0,-1){26}}
\put(70,28){\line(1,0){30}}

\put(70,28){\circle*{1}}
\put(60,28){\circle*{1}}
\put(30,28){\circle*{1}}
\put(100,28){\circle*{1}}
\put(30,32){\circle*{1}}
\put(100,32){\circle*{1}}
\put(60,2){\circle*{1}}
\put(70,2){\circle*{1}}

\put(27,32){\makebox{$0$}}
\put(103,32){\makebox{$\pi$}}
\put(65,35){\makebox{$X^{*}$}}
\put(103,26){\makebox{$\pi$}}
\put(27,26){\makebox{$0$}}
\put(72,0){\makebox{$0$}}
\put(56,0){\makebox{$\pi$}}
\put(72,23){\makebox{$\frac{\pi}{2}$}}
\put(56,23){\makebox{$\frac{\pi}{2}$}}
\put(50,15){\makebox{$\Psi^{*}$}}
\put(75,15){\makebox{$\Phi^{*}$}}

\end{picture}
so graphically we have shown (\ref{eq400:ps}).
\subsection{Half string approach}
\label{sec:HalfStringApproach}
For our purposes it is one of the easiest contructions, infact it uses heavily the possibility of describing the star product in terms of infinite dimensional matrices. This fact simplifies our job as we know how to define a $^{*}-$operation on a matrices algebra: it is the usual hermitian conjugation; given $\Phi\in \A$ then
\[(\Phi)^{*}=\Phi^{\dag}=(\overline{\Phi})^{t}.\]
It is also immediate the interpretation of the operations: the transposition consists in the exchange $x^{L}\leftrightarrow x^{R}$, so the $^{*}-$action on the field becomes

\begin{equation}\label{eq20:ps}
	(\Phi[x^{L},x^{R}])^{*}=\overline{\Phi[x^{R},x^{L}]}
\end{equation}
It is immediate to verify that
\begin{equation}
	(\Phi^{\dag})^{\dag}=\Phi~;~~(\Phi\circ \Psi)^{\dag}=\Psi^{\dag}\circ \Phi^{\dag}~;~~(\lambda\Phi)^{\dag}=\overline{\lambda}\Phi^{\dag}
\end{equation}
So we verified that $\A$ is indeed a $\st$algebra.\\
Moreover we can compare this approach with the previous one simply noticing that the transposition of the matrices given by the exchange $x^{L\mu}\leftrightarrow x^{R\mu}$ corresponds exactly to the inversion in the orientation $\sigma \rightarrow \pi-\sigma$. The complex conjugation is exactly the same.
\subsection{Moyal's approach}
\label{sec:MoyalApproach}
In this approch we can describe the string fields $\Phi[\left\{x_{2n-1}\right\},\left\{x_{2n}\right\}]$ as fields $\widetilde{\Phi}[\left\{x_{2n}\right\},\left\{p_{2n}\right\}]$ having performed a Fourier transform only on the odd coordinates $\left\{x_{2n-1}\right\}$.\\
So Witten's product between two fields $\Phi,\Psi\in\A$, becomes 
\[	(\Phi\W\Psi)\Rightarrow(\widetilde{\Phi}\M\widetilde{\Psi})=\]
\[=\widetilde{\Phi}[\left\{x_{2n}\right\},\left\{p_{2n}\right\}]\exp\Big(i\frac{\vartheta}{2}\sum^{\infty}_{n=1}\eta^{\mu\nu}\Big(\frac{\overleftarrow{\partial}}{\partial x^{\mu}_{2n}}\frac{\overrightarrow{\partial}}{\partial p^{\nu}_{2n}}-\frac{\overleftarrow{\partial}}{\partial p^{\nu}_{2n}}\frac{{\overrightarrow{\partial}}}{{\partial x^{\mu}_{2n}}}\Big)\Big)\widetilde{\Psi}[\left\{x_{2n}\right\},\left\{p_{2n}\right\}]
\]
Now if we define a $^{*}$ operation on the algebra $\A$ of functions $\widetilde{\Phi}$ as the usual complex conjugation (in the next formulas $c.c.$ stands for the usual complex conjugation) we fulfill all the requirements of a $\st$algebra:
\[^{*}:\A\longrightarrow \A\]
with $\widetilde{\Phi}^{*}:=\widetilde{\Phi}^{c.c.}$. Infact we have\\
\\
$\bullet$ $(\lambda\widetilde{\Phi})^{*}=\lambda^{c.c.}\widetilde{\Phi}^{c.c.}$;\\
\\
$\bullet$ $(\widetilde{\Phi}^{*})^{*}=\widetilde{\Phi}$;\\
\\
$\bullet$ $(\widetilde{\Phi}\M\widetilde{\Psi})^{*}=\widetilde{\Psi}^{*}\M\widetilde{\Phi}^{*}$ from
\[(\widetilde{\Phi}\M\widetilde{\Psi})^{*}=\]
\[=\widetilde{\Phi}[\left\{x_{2n}\right\},\left\{p_{2n}\right\}]^{c.c.}\exp\Big(-i\frac{\vartheta}{2}\sum^{\infty}_{n=1}\eta^{\mu\nu}\Big(\frac{\overleftarrow{\partial}}{\partial x^{\mu}_{2n}}\frac{\overrightarrow{\partial}}{\partial p^{\nu}_{2n}}-\frac{\overleftarrow{\partial}}{\partial p^{\nu}_{2n}}\frac{{\overrightarrow{\partial}}}{{\partial x^{\mu}_{2n}}}\Big)\Big)\widetilde{\Psi}[\left\{x_{2n}\right\},\left\{p_{2n}\right\}]^{c.c.}=\]
\[=\widetilde{\Psi}[\left\{x_{2n}\right\},\left\{p_{2n}\right\}]^{c.c.}\exp\Big(i\frac{\vartheta}{2}\sum^{\infty}_{n=1}\eta^{\mu\nu}\Big(\frac{\overleftarrow{\partial}}{\partial x^{\mu}_{2n}}\frac{\overrightarrow{\partial}}{\partial p^{\nu}_{2n}}-\frac{\overleftarrow{\partial}}{\partial p^{\nu}_{2n}}\frac{{\overrightarrow{\partial}}}{{\partial x^{\mu}_{2n}}}\Big)\Big)\widetilde{\Phi}[\left\{x_{2n}\right\},\left\{p_{2n}\right\}]^{c.c.}=\]
\[=\widetilde{\Psi}^{c.c.}\M\widetilde{\Phi}^{c.c.}=\widetilde{\Psi}^{*}\M\widetilde{\Phi}^{*}\]
So again we have the confirm that $\A$ is a $\st$algebra.

\section{Functional approach to $\A$}
\label{sec:FunctionalApproachToA}
The problem of finding a $\st$algebra was not too hard after all and we have seen that the whole of $\A$ is a $\st$algebra. Now we want to see if is it possible to introduce a $\Cs$-algebraic structure, and it is a harder challenge, expecially because we need to have a better defined mathematical background. At the end we will find that not the whole of $\A$ is a $\Cs$-algebra but only appropriate subsets. In particular we will find one of them.\\
In order to have a well defined mathematical problem we have to understand all the various  objects that enter the game.\\
\\
In this section we will show that it is possible to approach the task from a functional point of view in which the string field states $ \Phi$ become functionals over some suitable defined space.
\subsection{The space of string states $C_{X^{D}}(I)$}
\label{sec:TheSpaceOfStringStates}
The basic object we start with is the physical string $\xs$. It is described by a vector field defined on a closed interval $I=[0,\pi]$ and the only requirement is the continuity as we suppose the string does not break. So all possible string configurations are given by all possible continuous vector fields
\[x^{\mu}:I\longrightarrow X^{D}\]
where $D$ is the dimension of spacetime.\\
So we get a set 
\[C_{X^{D}}(I)=\left\{x^{\mu}:I\longrightarrow X^{D}~s.t.~x^{\mu}(\sigma)~continuous\right\}\]
Such vector fields can be described by their Fourier expansion
\begin{equation}\label{eq40:ps}
	x^{\mu}(\sigma)=x^{\mu}_{0}+\sqrt{2}\sum^{\infty}_{n=1}x^{\mu}_{n}\cos(n\sigma)
\end{equation}
where each $x_{n}\in\RR$ as the vector field is real: $\overline{x^{\mu}(\sigma)}=x^{\mu}(\sigma)$.\\
Now we can fix the attention on a single component $x^{\overline{\mu}}(\sigma)=x(\sigma)$. It is a continuous function on a closed interval $I$, so it is a $L^{2}(I)$ function, and, as shown in (\ref{eq40:ps}) it is written in components on a basis $\left\{e_{n}=\cos(n\sigma)\right\}$, so that it can be described by infinite coordinates $\left\{x_{n}\right\}^{\infty}_{n=0}$. They describe an infinite dimensional space, and if we suppose $x^{n}\in\RR~~\forall~n$, then they describe a $H=\RR^{\infty}$ manifold. This is an example of \emph{Hilbert manifold} that is a manifold that is locally homeomorphic to a Hilbert space (in our case we have a global homeomorphism).\\
So we have a map
\[\gamma:C_{X^{D}}(I)\longrightarrow H\]
which allows us to describe a generic string state using coordinates $\left\{x_{n}\right\}$ living in a subset of $H$ (and not all $H$ as the space $C_{X^{D}}(I)$ is not a Banach space in the $L^{2}(I)$ norm: there are $L^{2}(I)$ functions which are not continuous, like a step function on $I$). What is important here is the possibility of describing a string state by coordinates $\left\{x_{n}\right\}$ living in some subspace $U\subset H$.\\
So our starting point is a space $U$, and any string field $\Phi[x(\sigma)]=\langle x(\sigma)\vert \Phi\rangle$ can be defined as function on it:
\[\Phi:U\longrightarrow \C\]
as we suppose to have complex valued fields.\\
Now the string fields algebra $\A$ has a good mathematical foundation and we can consider the functional analysis on it, as we usually do for quantum mechanics where we deal with functions
\[\Psi:\RR^{d}\longrightarrow \C\]
and $d$ is a finite number. In that case we consider only appropriate subsets $\mathcal{G}\subset \mathcal{F}$ (where $\mathcal{F}$ is the space of all possible functions on $\RR^{d}$) with physical meaning (e.g. $\mathcal{G}=L^{2}$).\\
In this case too we can restrict the class of functions to appropriate subsets $\mathcal{U}\subset \A$ on which we can define a norm.
\section{Pre $\Cs$-algebras on finite dimensional spaces}
\label{sec:PreCsAlgebraOnFiniteDimensionalSpaces}
Looking at all various approaches we have to SFT, we see that mabye the most rigorous is Moyal's approach, at least if we consider a cutoff in the string modes, that is we consider all modes $\{x_{n}\}$ with $n\in [1,N]$ with $N$ finite. This implies that the space defined in the previous section is a finite dimensional manifold $\RR^{N}$. On such spaces we have a well defined functional analysis where instead of the pointwise product we use Moyal's product. Then only at the end we will take an inductive limit in order to extend the construction to all string modes, removing the cutoff.\\
The big problem we had to face was to find a space of functions that is a $\Cs$-algebra, or even more simply a pre $\Cs$-algebra in which the composition law is Moyal's product. The first hint in this direction was found in \cite{[17]}, while the mathematical details were found in \cite{[14]}.
\subsection{Construction and main properties of the space $\Sc({\RR^{2n}})$}
\label{sec:ConstructionAndMainPropertiesOfTheSpaceScRR2n}
We start with a finite even dimensional space $\RR^{2n}$ and the space $\Sc(\RR^{2n})$ of rapidly decreasing smooth functions, given by the conditions
\begin{equation}
	x^{k}\partial^{\alpha}f(x)~~~bounded~\forall~k\in\N;~\forall \alpha\in \N^{2n}
\end{equation}
Such space, called \emph{Schwartz space}, is a Frechet space, that is, its topology is given by a family of seminorms, either 
\begin{equation}\label{eq190:ps}
p_{a\gamma}(f)=\Vert x^{a}\partial^{\gamma}f\Vert_{\infty}=sup\vert x^{a}\partial^{\gamma}f\vert
\end{equation}
or
\begin{equation}\label{eq191:ps}
q_{a\gamma}(f)=\Vert x^{a}\partial^{\gamma}f\Vert_{1}
\end{equation}
and it is complete with respect to such families.\\
Now we want to see if $\Sc$ can be a $\Cs$-algebra (eventually a pre $\Cs$-algebra).\\
First of all we need to verify it is an algebra under Moyal's product.\\
Given two functions $f,g\in\Sc$ we define Moyal's product using (\ref{eq35:ps}) setting in particular $\vartheta=2$. Then we want to show that $f\star g\in\Sc$, and that Moyal's product is a continuous bilinear operation, that is we need to show that $x^{a}\partial^{\gamma}(f\star g)$ is bounded $\forall a\in\N; \forall \gamma \in \N^{2n}$.\\ 
So let us consider the following relations
\begin{equation}\label{eq200:ps}
	\partial_{i}(f\star g)=(\partial_{i}f)\star g+f\star \partial_{i}g
\end{equation}
\begin{equation}\label{eq201:ps}
	x_{i}(f\star g)=f\star x_{i}g+i\hat{\partial}_{i}f\star g=x_{i}f\star g-i f\star \hat{\partial}_{i}g
\end{equation}
where
\begin{displaymath}
	\hat{\partial}_{i}f:=\left\{ \begin{array}{ll}
	\partial_{i+n}f & \textrm{if $1\leq i \leq n$}\\
	-\partial_{i-n}f & \textrm{if $n<i\leq 2n$}
	\end{array} \right.
\end{displaymath}
Now, from the definition of Moyal's product we have that 
\begin{equation}
sup\vert f\star g\vert= \Vert f\star g\Vert_{\infty}	\leq \Vert f \Vert_{1}\Vert g \Vert_{1}
\end{equation}
so $f\star g$ is bounded. Now using (\ref{eq200:ps}) and (\ref{eq201:ps}) we find
\begin{equation}\label{eq205:ps}
	x^{a}\partial^{\gamma}(f\star g)=\sum_{b\leq a}\sum_{\epsilon\leq \gamma}(-i)^{\vert b\vert} \left( \begin{array}{c}
	a \\
	b
	\end{array} \right)\left(\begin{array}{c}
	\gamma \\
	\epsilon
	\end{array} \right) x^{a-b}\partial^{\gamma-\epsilon}f\star \hat{\partial}^{b}\partial^{\epsilon}g
\end{equation}
and using the definition of seminorms (\ref{eq190:ps}) and (\ref{eq191:ps}) we find
\begin{equation}\label{eq207:ps}
	p_{a\gamma}(f\star g)\leq \sum_{b\leq a}\sum_{\epsilon\leq \gamma} \left( \begin{array}{c}
	a \\
	b
	\end{array} \right)\left(\begin{array}{c}
	\gamma \\
	\epsilon
	\end{array} \right) q_{a-b,\gamma-\epsilon}(f)q_{0,\eta-\epsilon}(g)
\end{equation}
with $\eta_{j}=b_{j\pm n}$, so by induction, $f\star g\in \Sc$. Moreover (\ref{eq207:ps}) shows also that Moyal's product is continuous in the topology induced by the seminorms used.
\subsection{Operator norm on $\Sc(\RR^{2n})$}
\label{sec:OperatorNormOnScRR2n}
Now we can consider $\Sc(\RR^{2n})$ as a space of operators acting on a Hilbert space $\mathcal{H}=L^{2}_{\star}(\RR^{2n})$ where the norm on $L^{2}_{\star}$ is defined by means of  Moyal's product
\[\Vert~ \Vert_{2\star}:L^{2}_{\star}(\RR^{2n})\longrightarrow \RR^{+}\]
with 
\[\Vert f \Vert_{2\star}=\Big[\int d\mu~ \bar{f}\star f\Big]^{\frac{1}{2}}\]
It can be shown to be a norm.\\
So if we define the action of $\Sc$ on $L^{2}_{\star}(\RR^{2n})$ by means of a left $\star$ product:
\[L_{f}:L^{2}_{\star}(\RR^{2n})\longrightarrow L^{2}_{\star}(\RR^{2n})\]
with $g\rightarrow f\star g$.\\
We can wonder if the action of $\Sc(\RR^{2n})$ on $L^{2}_{\star}(\RR^{2n})$ still gives an element in $L^{2}_{\star}(\RR^{2n})$. The answer is yes, and it is a consequence of the theorem 8 of reference \cite{[14]} p.878:
\[\Vert f\star g\Vert_{2\star}\leq \Vert f \Vert_{2\star}\Vert g\Vert_{2\star}\]
In this way we obtain that elements of $\Sc(\RR^{2n})$ can be seen as bounded operators on the Hilbert space $L^{2}_{\star}(\RR^{2n})$.\\
Now we can define the $\Cs$-algebra norm for the operators $f\in \Sc$:
\begin{equation}
	\Vert f\Vert:=sup\Big(\Vert f\star g\Vert_{2\star}~~s.t.~~g\in L^{2}_{\star}(\RR^{2n})~;~ \Vert g\Vert_{2\star}\leq 1\Big)
\end{equation}
From the general theory of $\Cs$-algebra of bounded operators (see \cite{[18]}) and what said in section (\ref{sec2}) the space $\Sc(\RR^{2n})$ with such operator norm is a pre $\Cs$-algebra as it is not complete. Anyway it can be completed to obtain a $\Cs$-algebra.\\
\\
\textbf{Remark}: there are two different kinds of norm on the space $\Sc(\RR^{2n})$ and it is better to clarify this point in order to avoid confusion. As said at the beginning, as functional space it is a Frechet space, its topology is given by a family of seminorms, and it is complete with respect to such topology. Such seminorms are needed to show that $\Sc$ closes under Moyal product. \\
But $\Sc(\RR^{2n})$ can also be seen as a space of bounded operators acting on a "Moyal Hilbert space" $L^{2}_{\star}(\RR^{2n})$, and in this case we can define an operator norm, which has nothing to do with the previous family of seminorms, and with respect to which it is not complete.\\
\\
Let us notice that the squeezed states representing $D-$branes in VSFT, after we have used a cutoff on the string oscillator modes, are gaussians, so they are elements of $\Sc(\RR^{2n})$. This is good as the space we have found is meaningful for our purposes.
\section{Inductive limit}
\label{sec:InductiveLimit}
The construction seen so far is mathematically well posed, and satisfies our requirements, but it can be applied only after we have considered a cutoff procedure, so we are missing a part of the theory. The purpose of this section is to extend the above construction to the infinite dimensional case, and this can be done using a procedure which is quite general in the theory of $\Cs$-algebras, called \emph{inductive limit}.
\subsection{General construction}
\label{sec:GeneralConstruction}
First of all we notice that we have a pre $\Cs$-algebra $\Sc(\RR^{2n})$ for each positive $n$.\\
Now we describe the general procedure (see \cite{[18]} p.468), and later on we will specialize to our concrete case.\\
\\
We take a directed set $I$ \footnote{a directed set $I$ is a set with a partial ordering $\prec$ such that $\forall ~i,j\in I~\exists ~k\in I$ with $i\prec k,j\prec k$}. Then $\forall~ i\in I$ let $A_{i}$ be a $\Cs$-algebra; $\forall ~i,j\in I$ with $i\prec j$ let us consider the map 
\[F_{ji}:A_{i}\longrightarrow A_{j}~~~~;\]
then we require $F_{ji}$ to have the following properties:\\
\\
$\bullet$ $F_{ji}(a\circ b)=F_{ji}(a)\circ F_{ji}(b)$;\\
$\bullet$ $F_{ji}(a^{*})=(F_{ji}(a))^{*}$ (that is $F_{ji}$ is a $\st$homomorphism and this implies $\Vert F_{ji}(a)\Vert \leq \Vert a\Vert$);\\
$\bullet$ $F_{ii}=Id$\\
$\bullet$ $F_{kj}\circ F_{ji}=F_{ki}$ with $i\prec j\prec k$ and $i,j,k\in I$.\\
\\
A system $(A_{i},F_{ji})$ fulfilling all the requirements is called \emph{directed system of $\Cs$-algebras}.\\
Now we consider the set $P=\left\{<i,a>~s.t.~i\in I,~a \in A_{i}\right\}$ and let us introduce an equivalence relation $\sim$:
\begin{equation}
	<i,a>\sim <j,b>\Longleftrightarrow \lim_{k\rightarrow \infty}\Vert F_{ki}(a)-F_{kj}(b)\Vert=0
\end{equation}
So we can define equivalence classes $<i,a>^{\sim}$ and the set of all such classes is given by $\A'=P/\sim$. It can be shown that $\A'$ is a pre $\Cs$-algebra if we define the following operations:\\
\\
$\bullet$ $\alpha<i,a>^{\sim}=<i,\alpha a>^{\sim},~\forall~ \alpha\in \C$;\\
$\bullet$ $<i,a>^{\sim}+<j,b>^{\sim}=<k,F_{ki}(a)+F_{kj}(b)>^{\sim}$;\\
$\bullet$ $<i,a>^{\sim}\circ <j,b>^{\sim}=<k,F_{ki}(a)\circ F_{kj}(b)>^{\sim}$;\\
$\bullet$ $<i,a>^{\sim *}=<i,a^{*}>^{\sim}$;\\
$\bullet$ $\Vert <i,a>^{\sim}\Vert= \lim_{k\rightarrow \infty} \Vert F_{ki}(a)\Vert$.\\
\\
Notice that in the last line we have a well defined finite limit as $F_{ki}$ is a positive defined, norm decreasing homomorphism.\\
Of course the algebra $\A'$ can be completed to a $\Cs$-algebra.
\subsection{Application to the $\Sc(\RR^{2n})$ case}
\label{sec:ApplicationToTheScRR2nCase}
It is time to consider our case where $I=\N$, $A_{i}=\Sc(\RR^{2i})$, $A_{j}=\Sc(\RR^{2j})$ with $i<j$ and $2j-2i=2n$. Then let us put coordinates $(x_{1},\ldots,x_{2i})$ on $\RR^{2i}$ and $(x_{1},\ldots,x_{2i},y_{1},\ldots,y_{2n})$ on $\RR^{2j}$.\\
At this point let us consider the following maps:
\[F_{ji}:A_{i}\longrightarrow A_{j}\]
with
\begin{equation}\label{eq320:ps}
	F_{ji}(f(x_{1},\ldots,x_{2i}))=f(x_{1},\ldots,x_{2i})\exp\Big(-\sum^{2n}_{l=1}y^{2}_{l}\Big)
\end{equation}
where the product is the usual pointwise product.\\
If we have $F_{ii}:A_{i}\longrightarrow A_{i}$ then we define $F_{ii}(f)=f(x_{1},\ldots,x_{2i})\times 1$.\\
Then, by construction
\[F_{kj}F_{ji}=F_{ki}\]
and 
\[F(f^{*})=(F(f))^{*}.\]
So the hardest check is
\begin{equation}\label{eq305:ps}
	F_{ji}(f\star g)=F_{ji}(f)\star F_{ji}(g)
\end{equation}
In order to show it let us consider the following functions
\[F(f)=f'(\vec{x},\vec{y})=f(\vec{x})\exp\Big(-\sum^{2n}_{l=1}y^{2}_{l}\Big)\]
\[F(g)=g'(\vec{x},\vec{y})=g(\vec{x})\exp\Big(-\sum^{2n}_{l=1}y^{2}_{l}\Big)\]
Then
\[(f'\star g')(\vec{x}'',\vec{y}'')=(f'\star g')(\vec{X}'')=\]
\[=\int f'(\vec{X}')g'(\vec{X})\exp[i(\vec{X}''\Lambda \vec{X}+\vec{X}\Lambda \vec{X}'+\vec{X}'\Lambda \vec{X}'')]d\vec{X}' d\vec{X}\]
but we can easily factorize the exponents:
\begin{equation}\label{eq300:ps}
\vec{X}''\Lambda \vec{X}=\vec{x}''\Lambda_{a}\vec{x}+\vec{y}''\Lambda_{b} \vec{y}
\end{equation}
the same for the other two bits, so plugging in the above relation we find
\[(f'\star g')(\vec{x}'',\vec{y}'')=\int f(\vec{x})\exp\Big(-\sum^{2n}_{l=1}y^{2}_{l}\Big)g(\vec{x}')\exp\Big(-\sum^{2n}_{l=1}y'^{2}_{l}\Big)\times\]
\[\times\exp[i(\vec{x}''\Lambda_{a}\vec{x}+\vec{y}''\Lambda_{b}\vec{y}+\vec{x}\Lambda_{a}\vec{x}'+\vec{y}\Lambda_{b}\vec{y}'+\vec{x}'\Lambda_{a}\vec{x}''+\vec{y}'\Lambda_{b}\vec{y}'')]d\vec{x} d\vec{x}' d\vec{y} d\vec{y}'=\]
\[=\int f(\vec{x})g(\vec{x}')\exp[i(\vec{x}''\Lambda_{a}\vec{x}+\vec{x}\Lambda_{a}\vec{x}'+\vec{x}'\Lambda_{a}\vec{x}'')]d\vec{x}d\vec{x}'\times\]
\[\times \int \exp\Big(-\sum^{2n}_{l=1}y^{2}_{l}\Big)\exp\Big(-\sum^{2n}_{l=1}y'^{2}_{l}\Big)\exp[i(\vec{y}''\Lambda_{b}\vec{y}+\vec{y}\Lambda_{b}\vec{y}'+\vec{y}'\Lambda_{b}\vec{y}'')]d\vec{y}d\vec{y}'=\]
\[=(f\star g)(\vec{x}'')\times(G\star G)(\vec{y}'')\]
where we defined with $G$ the gaussian function such that under Moyal's product is a projector: $G\star G=G$. So we can show also relation (\ref{eq305:ps}) infact we have:
\[F_{ji}(f\star g)=(f\star g)(x_{1},\ldots,x_{2i})\exp\Big(-\sum^{2n}_{l=1}y^{2}_{l}\Big)\]
while
\[F_{ji}(f)\star F_{ji}(g)=\Big(f(x_{1},\ldots,x_{2i})\exp\Big(-\sum^{2n}_{l=1}y^{2}_{l}\Big)\Big)\star\Big(g(x_{1},\ldots,x_{2i})\exp\Big(-\sum^{2n}_{l=1}y^{2}_{l}\Big)\Big)=\]
\[=[(f\star g)(x_{1},\ldots,x_{2i})]\times [G\star G](y_{1},\ldots,y_{2n})=\]
\[=(f\star g)(x_{1},\ldots,x_{2i})\times G(y_{1},\ldots,y_{2n})\]
so we satisfied the requirements.\\
So, summarizing, we managed to define $\st$homomorphisms $F_{ji}$ and so from the general procedure seen before, we manage to build a pre $\Cs$-algebra $\mathfrak{S}\subset \A$.
\section{Conclusions}
\label{sec:Conclusions}
The algebra $\mathfrak{S}$ we have built is very interesting for our purposes. First of all it is possible to consider generalizations of the maps (\ref{eq320:ps}) \footnote{for example we can multiply a generic function $f(\vec{x})\in \Sc(\RR^{2i})$ by $P(\vec{y})G(\vec{y})$ where $P(\vec{y})$ is a suitable polynomial and $G(\vec{y})$ is the usual Gaussian.}, so that $\mathfrak{S}$ contains all various states found in VSFT, which represent nonperturbative solutions of the VSFT equation of motion (in the matter sector):
\[\Phi\W\Phi=\Phi,\]
physically interpreted as $D$-branes and which are clearly projectors.\\
Moreover now that we have a mathematical description of string states we could try to find all projectors. The problem is that we do not have the completion of algebras $\Sc(\RR^{2i})$, so when we take the inductive limit if we take only the pre $\Cs$-algebra without paying attention to the completion it can happen that we miss some projectors.\\
Anyway, if we managed to get the completion we would have in principle all possible projectors and it would be very interesting, at least for three reasons:\\
\\
$\bullet$ we would get all $VSFT$ solutions and this would be important for the analysis of the non perturbative behaviour of string field theory;\\
$\bullet$ the knowledge of all projectors allow us to build the $K$-theory of $\A$. As already said, the $K$-theory is useful to check the instability of the $D$-brane states found. This would be a further confirmation that the states found in VSFT are indeed $D$-branes;\\
$\bullet$ we could try to extend in some way such analysis to the superstring field theory and see if the already known results about the $RR$ charges of $D$-branes can also be obtained in this formalism.\\
\\
We leave these interesting challenges for future work.
\\
\\
\large\textbf{Acknowledgments}\\
\\
\normalsize The author is very grateful to G. Landi, F. Lizzi and U. Bruzzo for the very helpful discussions and correspondence. A particular ackowledgment to professor L. Bonora for the reading of the work and very useful comments and suggestions. 
\end{document}